%% file: main.tex
\documentclass[conference]{IEEEtran}
\IEEEoverridecommandlockouts
\usepackage{cite}
\usepackage{amsmath,amssymb,amsfonts,bm}
\usepackage{tabularx}
\usepackage{booktabs}
\usepackage{algorithm,algpseudocode}
\usepackage{graphicx}
\usepackage{textcomp}
\usepackage{xcolor}
\def\BibTeX{{\rm B\kern-.05em{\sc i\kern-.025em b}\kern-.08em
    T\kern-.1667em\lower.7ex\hbox{E}\kern-.125emX}}
    
\usepackage{float}

\input{syntax}  

\begin{document}

\title{Spectral Efficiency Optimization for mmWave Wideband MIMO RIS-assisted Communication \\
\thanks{This material is based upon work supported by the NSF GRFP under Grant No. DGE-1610403}}

\author{\IEEEauthorblockN{Pooja Nuti, Elyes Balti and Brian L. Evans}
\IEEEauthorblockA{\textit{Wireless Networking and Communications Group}\\
The University of Texas at Austin, Austin, TX USA\\
pnuti@utexas.edu, ebalti@utexas.edu, bevans@ece.utexas.edu}
}

\maketitle

\begin{abstract}
Reconfigurable Intelligent Surfaces (RIS) are passive or semi-passive heterogeneous metasurfaces and consist of many tunable elements. RIS is gaining momentum as a promising new technology to enable transforming the propagation environment into controllable parameters. In this paper, we investigate the co-design of per-subcarrier power allocation matrices and multielement RIS phase shifts in downlink wideband MIMO transmission using 28 GHz frequency bands. Our contributions in improving RIS-aided links include (1) enhanced system modeling with pathloss and blockage modeling, and uniform rectangular array (URA) design, (2) design of gradient ascent co-design algorithm, and (3) asymptotic (Big O) complexity analysis of proposed algorithm and runtime complexity evaluation.
\end{abstract}


\input{introduction}

\input{systemmodel}

\input{design}
\input{proposed_solution}
\input{numericalresult}
\input{computational_complexity}

\input{conclusion}

\bibliographystyle{IEEEtran}
\bibliography{references}
\end{document}

%% file: syntax.tex
\def\ba{{\mathbf{a}}}

\def\bee{{\mathbf{e}}}

\def\bp{{\mathbf{p}}}

\def\bv{{\mathbf{v}}}

\def\bx{{\mathbf{x}}}
\def\by{{\mathbf{y}}}

\def\b0{{\mathbf{0}}}

\def\bA{{\mathbf{A}}}

\def\bH{{\mathbf{H}}}
\def\bI{{\mathbf{I}}}

\def\bQ{{\mathbf{Q}}}

\def\bU{{\mathbf{U}}}

\def\bX{{\mathbf{X}}}
\def\bY{{\mathbf{Y}}}
\def\bZ{{\mathbf{Z}}}

\def\sfj{{\mathsf{j}}}

\def\sf0{{\mathsf{0}}}  

\DeclareMathOperator{\diag}{diag}

%% file: introduction.tex
\section{Introduction}
\label{section:introduction}
The advent of fifth-generation (5G) communication systems promises rapid increases in connectivity in the coming years. A primary enabling technology of 5G communication is the use of millimeter wave (mmWave) frequency bands and beyond (30--300GHz) in order to meet increasing demands of communication systems and circumvent the limited spectrum resources in the sub-6 GHz band \cite{mmWave_main_ref}. Enabling technologies for mmWave communication systems include massive multiple-input multiple-output (MIMO), millimeter wave (mmWave) communication systems and ultra-dense networks (UDNs) which can offer large bandwidths in conjunction with greater beamforming and spatial-multiplexing gains from antenna arrays to meet demands for higher data rates and support a growing number of mobile devices \cite{capacity_characterization_MIMO,federated_learning_ris}.

The mmWave channel is susceptible to blockage, which can occur due to objects in the environment, and can suffer from severe pathloss under omnidirectional antennas, which can be compensated by highly directional antennas and appropriate beamforming techniques \cite{mmWave_main_ref}. However, many proposed solutions within mmWave frequency bands lead to increased hardware requirements and power consumption. For instance, UDNs are proposed to deploy dense base stations (BS) or hotspots which will lead to higher hardware cost and power consumption. Massive MIMO, although large antenna arrays can improve communication links, requires complex signal processing and increased number of radio frequency (RF) chains, which is turn leads to added hardware costs \cite{federated_learning_ris}. Consequently, there is need for spectral-efficient, energy-efficient, and cost-effective solutions for future wireless networks.
\begin{figure}[t!]
    \centering
    \includegraphics[width=\linewidth]{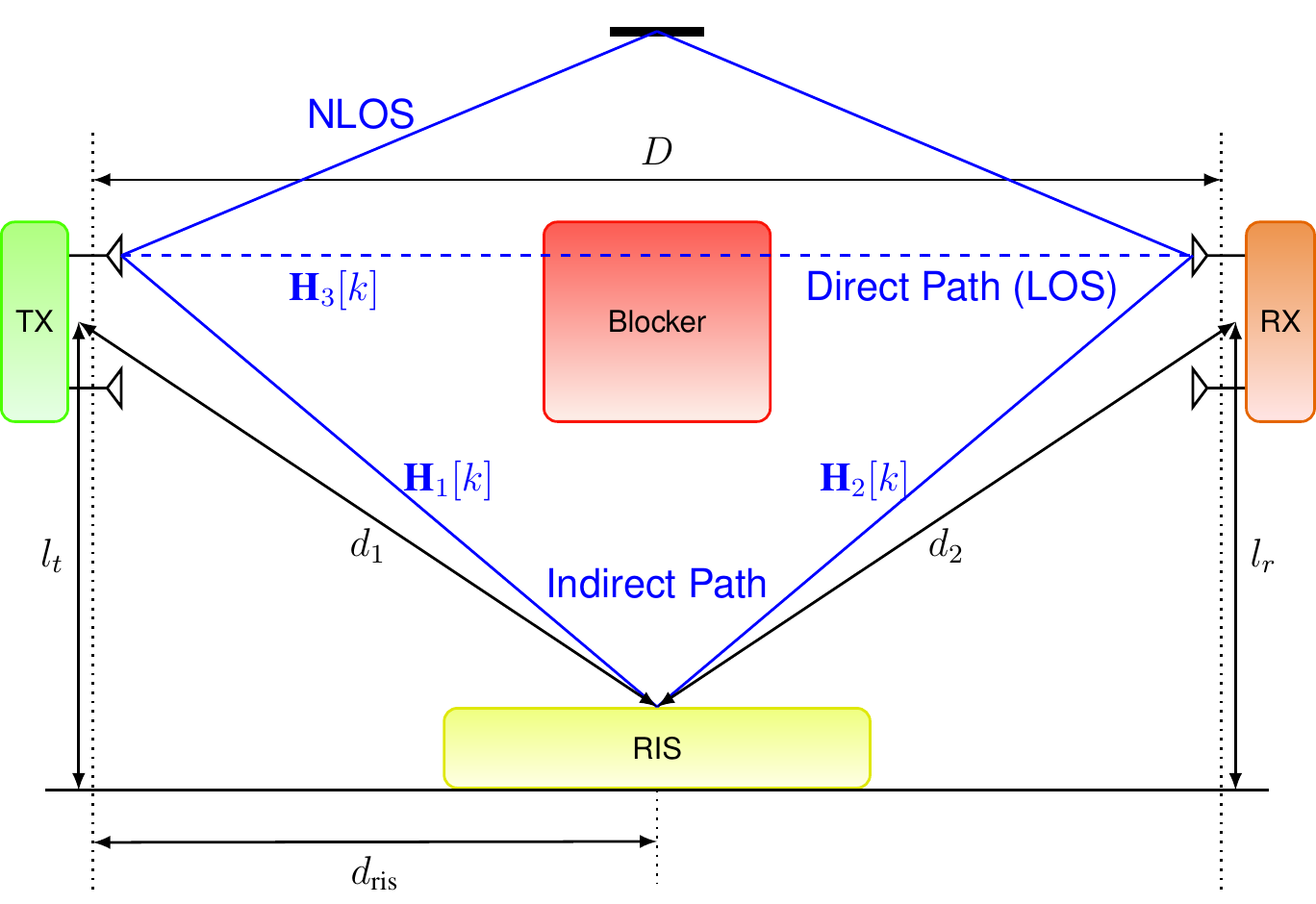}
    \caption{Aerial view of the proposed system.  
    A basestation transmits to a user equipment along a direct path and an indirect path.  The direct path consists of line-of-sight (LOS) and non-line-of-sight (NLOS) paths, and the indirect path reflects off of the reconfigurable intelligent system (RIS).}
    \label{fig:system_model}
    \vspace*{-0.1in}
\end{figure}

Recently, the concept of reconfigurable intelligent surfaces (RISs) has been gaining traction as an emerging solution for mitigating some of the high energy consumption in future communication systems. A primary enabling technology considered for RISs are metasurfaces. Metasurfaces are thin panels comprised of a large number of low-cost elements used to supplement existing communication systems. RISs are intended to operate with low cost and power demands, and as such, the elements are often assumed passive; i.e., they neither transmit nor receive \cite{intelligent_walls_smart_environment_paper, metasurface_magazine, overview_paper}. The reflection coefficients and phases of the RIS can be tuned independently by means of smart switches and a RIS controller. A RIS can be designed so that it reflects signals (i) constructively with those from the other paths to enhance the desired signal power, or (ii) destructively to cancel interference \cite{overview_RIS_journal, overview_RIS_magazine, overview_paper, interference_cancel,pooja_miso}. In this work we consider the mmWave MIMO orthogonal frequency division multiplexing (OFDM) RIS-assisted system in Fig. \ref{fig:system_model}.

A substantial amount of prior work regarding RIS pertains to a single-input single-output (SISO) or multiple-input single-output (MISO) setting \cite{capacity_characterization_MIMO} in sub-6 GHz systems. In \cite{joint_reflect_precode_ris}, a narrowband MIMO RIS-assisted system is considered in which the authors jointly optimize the precoder and RIS phase shifts by minimizing the symbol error rate. In \cite{capacity_characterization_MIMO}, the authors propose an alternating optimization (AO) method of the RIS phase shifts and power allocation matrix in both narrowband and wideband MIMO systems. In \cite{mimo_ofdm_protocol}, the authors propose a practical transmission protocol using pilot training to estimate the channels in a MIMO-OFDM system, and then propose an AO algorithm to optimize the RIS phase shifts and power allocation matrix. The problem of mutual coupling in an SISO-OFDM setting is considered in \cite{binary_ris}, in which the authors propose a low-complexity solution for channel estimation and RIS configuration including practical reflection amplitudes and binary RIS phase shifts. A mmWave MIMO communication system is considered in \cite{semipassive_ris_ce} in which a semi-passive RIS is considered with low-resolution active elements in the RIS for channel estimation. We extend upon the work in \cite{achievable_rate_opt_MIMO} which configures a RIS by projected gradient ascent (PGA) upon the RIS phase shifts and power allocation matrix in a MIMO narrowband sub-6 GHz setting. Gradient ascent achieves the same performance as the AO method proposed in \cite{achievable_rate_opt_MIMO} with lower computational complexity in the studied settings.

We distinguish this work from prior work by considering the joint optimization of the RIS phase shifts and power allocation matrices in a mmWave MIMO-OFDM downlink communication system between a BS and single user equipment (UE) with practical environment considerations. In a substantial amount of prior work, pathloss is not modeled and the direct path between a BS and UE is assumed to always be available. A significant use case of RIS is signal focusing at a UE when the direct path is unavailable. This becomes more important in mmWave frequencies and beyond. We adapt the RIS pathloss in \cite{achievable_rate_opt_MIMO} to the mmWave setting with directional antennas and model blockage on the direct path between the BS and UE. 

The rest of this paper is organized as follows. The system model is discussed in Section \ref{section:system_model} and the optimization problem is written in Section \ref{section:prob_form}. In Section \ref{section:proposed_solution} we introduce the proposed PGA method considered for RIS configuration and power allocation method considered. In Section \ref{section:numerical_results}, we show through simulation the performance of the proposed optimization in a RIS-assisted system over a non-optimized RIS-assisted system and system without a RIS in various settings. Lastly, in the Section \ref{section:complexity} present the a complexity analysis of the proposed algorithms using Big-O analysis and runtime evaluation. The paper is concluded in Section \ref{section:conclusion}.

%% file: systemmodel.tex
\section{System Model}
\label{section:system_model}
We consider a mmWave MIMO-OFDM RIS-assisted downlink communication system consisting of a direct path between a basestation (BS) and single-user equipment (UE) and indirect path between the BS and UE through a single RIS as shown in Fig. \ref{fig:system_model}. In Fig. \ref{fig:system_model}, the direct path consists of a blocker which we use to investigate blockage effects on spectral efficiency. The direct path consists of both a line-of-sight (LOS) and non-line-of-sight (NLOS) path which is considered in the case of blockage of the direct LOS path. The transmitter is equipped with $N_\text{t}$ antennas to transmit $N_\text{s}$ streams and the receiver consists of $N_\text{r}$ antennas. Both the BS and UE are equipped with uniform rectangular arrays (URA). The RIS consists of $N_\text{RIS}$ passive elements with tunable phase shifts $[0,2\pi]$, which are assigned via a RIS controller. In this work, we consider unit reflection coefficients on each of the RIS elements in order to ensure the RIS is passive and does not amplify the signal.

\subsection{Channel Model}
We consider Rician fading channels with Rician factor $\tilde{K}$ 
\begin{equation}
\textsf{H}_i[l] = \sqrt{\frac{\tilde{K}}{\tilde{K}+1}}\bar{\textsf{H}}_i[l] + \sqrt{\frac{1}{\tilde{K}+1}}\tilde{\textsf{H}}_i[l]
\label{freq_selective_rician_channels}
\end{equation}
where $i = 1,2,3$ denotes the channel index and $l$ is the index corresponding to a delay tap of the time-domain channel.
We denote $\bH_1[k] \in \mathbb{C}^{N_\text{RIS} \times N_\text{t}}$ as the channel between the BS and RIS indexed by the $k$-th subcarrier, $\bH_2[k] \in \mathbb{C}^{N_\text{r} \times N_\text{RIS}}$ as the channel between the RIS and UE indexed by the $k$-th subcarrier, and $\bH_3[k] \in \mathbb{C}^{N_\text{r} \times N_\text{t}}$ as the direct channel between the BS and UE indexed by the $k$-th subcarrier. The number of taps in the three channels are denoted as $L_1, L_2, L_3$. The entries of $\tilde{\textsf{H}}_i[l]$ are independently and identically distributed (i.i.d.) according to $\mathcal{CN}(0,1)$. The channels $\bar{\textsf{H}}_i[l]$ are defined by the geometry of the antenna arrays and environment. In particular, the following channel is adopted \cite{mmwave_geometric_channel_model}
\begin{equation}
\bar{\textsf{H}}_i[l] = \sqrt{\frac{N_\text{RX} N_\text{TX}}{RC}} \sum_{c=0}^{C-1}\sum_{r=0}^{R-1}\beta_{rc} \ba_{\text{r}}(\phi_{c,r}^{\text{r}},\varphi_{c,r}^{\text{r}})\ba_{\text{t}}^*(\phi_{c,r}^{\text{t}},\varphi_{c,r}^{\text{t}})
\label{geometric_channel}
\end{equation}
where $\beta_{rc}$ is the complex gain of the $r$-th ray in the $c$-th cluster, $C$ is the number of clusters, $R$ is the number of rays per cluster, $(\phi_{c,r}^\text{r},\varphi_{c,r}^\text{r})$ are the azimuth (elevation) angles of arrival. Similarly, $(\phi_{c,r}^\text{t},\varphi_{c,r}^\text{t})$ are the azimuth (elevation) angles of departure. We denote the array response vectors of the BS and UE URAs as $\ba_{\text{r}}(\phi_{c,r}^{\text{r}},\varphi_{c,r}^{\text{r}})$ and $\ba_{\text{t}}^*(\phi_{c,r}^{\text{t}},\varphi_{c,r}^{\text{t}})$.

\subsection{Blockage and Pathloss Modeling}
In the frequency domain, the frequency-selective equivalent channel matrix is the sum of the direct and indirect paths
\begin{equation}
    \bH_{\text{eq}}[k] = \sqrt{\rho_{\text{direct}}} \bH_3[k] + \sqrt{\rho_{\text{indirect}}} \bH_2[k] \bm \Phi \bH_1[k],   \forall k
    \label{equivalent_channel}
\end{equation}
where $k$ is the subcarrier index, $\rho_{\text{direct}}$ and $\rho_{\text{indirect}}$ are the pathlosses for direct and indirect paths, respectively. In (\ref{equivalent_channel}), the phase shifts of the RIS elements are arranged in the diagonal matrix $\bm \Phi \in \mathbb{C}^{N_\text{RIS} \times N_\text{RIS}}$, $\bm\Phi=\text{diag}\{ e^{\sfj \phi_1}, \ldots, e^{\sfj \phi_{N_\text{RIS}}} \}$. Note that $\bm \Phi$ is common across subcarriers $k$ in (\ref{equivalent_channel}) , due to the passive nature of the RIS not including baseband processing \cite{capacity_characterization_MIMO}. In \cite{achievable_rate_opt_MIMO}, the authors provide pathloss expressions for the direct and indirect links in Fig. \ref{fig:system_model} for a sub-6 GHz system, we update these pathloss expressions to include the antenna gain associated with directional antennas necessary for communication in mmWave settings\cite{mmWave_main_ref}.
We update the indirect link pathloss from \cite{achievable_rate_opt_MIMO} as follows:
\begin{equation}
   \rho_\text{indirect} = \frac{256G_\text{t}G_\text{r}\pi^2d_1^2d_2^2}{\lambda^4(l_\text{t}/d_1 + l_\text{r}/d_2)^2}
    \label{indirect_pathloss}
\end{equation}
In \cite{achievable_rate_opt_MIMO}, the distance between the BS array center and RIS center is given as $d_1 = \sqrt{d_\text{RIS}^2 + l_\text{t}^2}$, the distance between the RIS and UE array centers is $d_2 = \sqrt{(D-d_\text{RIS})^2 + l_\text{r}^2}$, and $G_\text{t}, G_\text{r}$ are the antenna gains for the BS and UE URAs. 

An important aim of RIS is to assist communication in the case of blockage in the direct path between a BS and UE. As such, we extend from prior work such as \cite{achievable_rate_opt_MIMO} to consider a more practical scenario in which the direct path between the BS and UE may be blocked, this is especially relevant in a mmWave setting. We model blockage of the direct path through the probability of the direct path being LOS, we denote a threshold probability $P_\text{LOS}$ as \cite{plos_ref} 
\begin{equation}
   P_\text{LOS} = e^{-(\sqrt{D^2 + (l_\text{t} - l_\text{r})^2} -10)/50}
\end{equation}
Letting $K_0 = \left(\lambda/4\pi d_0\right)^2G_\text{t}G_\text{r}$, we adopt a Bernoulli pathloss model for the direct path between the BS and UE as follows:
\begin{equation}
  \rho_\text{direct} =
  \begin{cases}
                                   K_0\left(\frac{d_0}{\sqrt{D^2 + (l_\text{t} - l_\text{r})^2}}\right)^{\alpha_{\text{LOS}}}&  \text{if} ~\text{LOS} \\
                                   K_0\left(\frac{d_0}{\sqrt{D^2 + (l_\text{t} - l_\text{r})^2}}\right)^{\alpha_{\text{NLOS}}}& \text{if} ~\text{NLOS}
   \end{cases}
   \label{prob_los_model}
\end{equation}
where $d_0$ is the reference distance assumed to be 1, $\alpha_\text{DIR}$ is the pathloss exponent for the direct path. The distances in the system are indicated in Fig. \ref{fig:system_model}, where $D$ is the distance between the BS and UE, $l_\text{t}$ is the height of the BS, $l_\text{r}$ is the height of the UE. The distance between the center of the RIS and the center of the BS antenna array is denoted as $d_\text{ris}$. In the case that the direct path is in blockage in (\ref{prob_los_model}), a different pathloss exponent $\alpha_\text{NLOS}$ and the number of clusters and rays per clusters is set to higher than those of the case where the direct path is not in blockage. This setting further accounts for the increased scattering observed in the NLOS path in Fig. \ref{fig:system_model}. 
We can express the received signal $\by[k] \in \mathbb{C}^{N_{\text{r}} \times 1}$ as
\begin{equation}
\by[k] = \bH_{\text{eq}}[k]\bx[k] + \bv[k]
\label{frequency_domain_received_signal}
\end{equation}
where $\bv[k]$ is the noise vector on the $k$-th subcarrier, which is distributed according to $\mathcal{CN}(0,\sigma_\text{n}^2\bI)$.

%% file: design.tex
\section{Problem Formulation}
\label{section:prob_form}
In this work, we consider the problem of maximizing the spectral efficiency at the UE by jointly optimizing the transmit covariance matrices $\bQ[k]$ according to power constraint in \eqref{optimization_func} for each subcarrier and the $\bm\Phi$.  The rate optimization for the considered system with perfect CSI can be stated as follows:
\begin{equation}
\begin{aligned}
\max_{\bm \Phi, \{\bQ[k]\}_{k=1}^K}  \quad & \frac{1}{K}\sum_{k=1}^K \log_2 \det \left(\bI_{N_\text{r}} + \frac{1}{\sigma_\text{n}^2} \bH_{\text{eq}}[k] \bQ[k] \bH_{\text{eq}}[k]^\text{H}\right)\\
\textrm{s.t.} \quad &\left|\bee_i^T \bm \Phi \bee_i \right| = 1, \quad  i = 1 \ldots N_\text{RIS}\\
  &\sum_\text{k=1}^K \sum_\text{g=1}^{N_\text{s}} P_{k,g} \leq P_\text{t}    \\
  &\bQ[k] \succeq \bm 0, k = 1, \ldots, K 
\end{aligned}
\label{optimization_func}
\end{equation}
$\bee_i$ is the $i$-th standard Euclidean basis vector of size $N_{\text{RIS}}$.

%% file: proposed_solution.tex
\section{Proposed Solution}
\label{section:proposed_solution}
\subsection{RIS Phase Optimization: PGA Algorithm}
In order to define the PGA method, the gradient on the objective function in the optimization problem in \eqref{optimization_func} with respect to $\bm \Phi$ is required.  Mathematically, we can write the desired gradient as
\begin{equation}
 \frac{\partial\sum_{k=1}^k \log_2 \det (\bI_{N_\text{r}} + \frac{1}{\sigma_\text{n}^2} \bH_{\text{eq}}[k] \bQ[k] \bH_{\text{eq}}[k]^\text{H})}{\partial \bm \Phi}
\label{desired_derivative}
\end{equation}
Note that $\bm \Phi$ does not vary with subcarrier index $k$, therefore the gradient in \eqref{desired_derivative} evaluates to the sum of the gradients obtained by each subcarrier, therefore we show the gradient for a particular subcarrier. 
We let $\bA_{k}(\bm \Phi) = (\bI_{N_\text{r}} + \frac{1}{\sigma_\text{n}^2} \bH_{\text{eq}}[k] \bQ[k] \bH_{\text{eq}}[k]^\text{H})$ and rewrite the gradient from \eqref{desired_derivative} as
\begin{equation}
 \frac{\sum_{k=1}^k \partial \log_2 \det (\bA_k(\bm \Phi)}{\partial \bm\Phi}
\label{simplified_derivative}
\end{equation}
We note that the gradient in \eqref{simplified_derivative} is a matrix containing the derivative of the scalar $\log_2\det(\bA_{k}(\bm\Phi))$ with respect to each of the elements in $\bm\Phi$. We choose to find the derivatives with respect to each of the diagonal elements of $\bm \Phi$ and then build the overall gradient matrix for the given subcarrier.
\begin{equation}
\frac{\partial\log_2\det(\bA_k(\bm\Phi))}{\partial \bm\Phi_{i,j}} = \text{Trace}\left[\bA_k(\bm\Phi)^{-1}\frac{\partial\bA_k(\bm\Phi)}{\partial \bm\Phi_{i,j}}\right]
\label{overall_deriv_expression}
\end{equation}
We now define the derivative of $\partial \bA_{k}(\bm\Phi)$ with respect to an element $\bm\Phi_{i,j}$.  Let $\bX_k = \frac{1}{\sigma_\text{n}^2}\bH_2[k]$, $\bY_k = \bH_1[k]\bQ[k]\bH_3[k]^\text{H}$, and $\bZ_k = \bH_1[k]\bQ[k]\bH_1[k]^\text{H}\bm\Phi^\text{H}\bH_2[k]^\text{H}$. 
\begin{equation}
\frac{\partial\bA_k(\bm\Phi)}{\partial \bm\Phi_{i,j}} =  \frac{\partial(\bX_k\bm\Phi\bY_k + \bX_k\bm\Phi\bZ_k)}{\partial \bm\Phi_{i,j}}
\label{matrix_product_deriv}
\end{equation}
The gradient in \eqref{matrix_product_deriv} evaluates the gradient for both the diagonal or off-diagonal elements of $\bm\Phi$. Since we are interested in the gradient with respect to only the diagonal RIS elements, we further simplify the expression in \eqref{matrix_product_deriv} in order to evaluate the gradient for the diagonal elements of $\bm\Phi$
\begin{equation}
\frac{\partial\bA_k(\bm\Phi)}{\partial \bm\Phi_{i,i}} = \bX_k \diag\left({\bee_i}\right) (\bY_k + \bZ_k)
\label{simplified_grad_part}
\end{equation}
Substituting \eqref{simplified_grad_part} into \eqref{overall_deriv_expression} leads to the desired gradient over $N_\text{RIS}$ elements. 

In order to enforce the unit modulus constraints on $\{\phi\}_{i=1}^{N_\text{RIS}}$, after updating the configuration of $\{\phi\}_{i=1}^{N_\text{RIS}}$ we project the solution back into the space of feasible solutions through the projection
\begin{equation}
\mathcal{P}(\phi_i) = \frac{\phi_i}{|\phi_i|}, \  i = 1 \ldots N_\text{RIS}
\label{projection}
\end{equation}
\subsection{Power Allocation: Spatial-Frequency Waterfilling}
Assuming CSI at the BS is available, the transmit covariance matrices $\bQ[k]$ should be optimized. In \cite{capacity_characterization_MIMO}, CVX is used, while in \cite{achievable_rate_opt_MIMO} one-dimensional waterfilling with a PGA update on the transmit covariance matrix in the narrowband sub-6 GHz MIMO case considered. We consider spatial-frequency waterfilling for designing $\bQ[k]$. We consider the power constraint in \eqref{optimization_func} and the method presented in \cite{sf_waterfilling_main} in which power is allocated across both streams and subcarriers. In \cite{sf_waterfilling_main}, the authors discuss that in the case of frequency-selective MIMO systems with CSI available at the transmitter, spactial-frequency waterfilling performs better than single dimensional waterfilling either in the frequency domain or in the spatial domain. The power allocation across subcarriers and streams $P_{k,g}$ in \eqref{optimization_func} is computed by
\begin{equation}
P_{k,g} = \max\left(0,\frac{1}{\lambda_\text{sf}} - \frac{1}{\lambda_{k,g}}\right)
\label{power_values}
\end{equation}
where $\lambda_\text{sf}$ is the cutoff frequency, which is iteratively updated and computed using the power constraint in \eqref{optimization_func} and \eqref{power_values} as defined in \cite{sf_waterfilling_main}. Spatial-frequency waterfilling assigns more power to transmission streams with larger power gain. 
The joint optimization of $\bm\Phi$ and $\bQ[k]$ is outlined in Alg. \ref{alg:grad_method_se}.
\begin{algorithm}[H]
\caption{Projected Gradient Ascent optimizing ${\cal R}(\bQ[k],\bm \Phi)$}\label{alg:grad_method_se}
\begin{algorithmic}[1]
\State \textbf{Initialize $\mathbf{\Phi}$ with random phases taken from $\mathcal{U}[0,2\pi]$}
\State \textbf{Set} $n = 0$
\State \textbf{Evaluate ${\cal R}_n(\bQ[k],\bm \Phi)$ using $\bQ[k]$, $\bm\Phi$} 
\While {$|{\cal R}_{n+1}(\bQ[k],\bm \Phi) - {\cal R}_{n}(\bQ[k],\bm \Phi)| \geq \epsilon$}
\vspace*{.15mm}
\State \textbf{Calculate  $\nabla_{\mathbf{\Phi_{n}}} {\cal R}(\bQ[k],\bm \Phi)$ } 
\vspace*{.15mm}
\State \textbf{Update $\diag(\bm \Phi_{n+1})$ as:}
\vspace*{.15mm}
\State \hspace{1mm} $\diag(\bm\Phi_{n+1}) = \diag(\bm\Phi_n) + \mu\diag(\nabla_{\bm{\Phi}_{n}} {\cal R}(\bQ[k],\bm \Phi))$
\State \textbf{Project phase shifts in $\mathbf{\Phi}_{n+1}$ back to feasible set } 
\vspace{.15mm}
\State \textbf{Update each transmit covariance matrix $\bQ_{n+1}[k]$}
\vspace*{.15mm}
\State \textbf{Evaluate ${\cal R}_{n+1}(\bQ[k],\bm \Phi)$ using $\bQ_{n+1}[k]$, $\bm\Phi_{n+1}$} 
\vspace*{.15mm}
\State \textbf{if  ${\cal R}_{n+1}(\bQ[k],\bm \Phi) \leq {\cal R}_{n}(\bQ[k],\bm \Phi) $} \\
 \hspace*{.25in} \textbf{Update learning rate $\mu$ as $\mu = \mu/10$} 
\EndWhile
\State \textbf{Evaluate the optimized ${\cal R}(\bQ[k]^\star,\bm \Phi^\star)$ } 
\end{algorithmic}
\end{algorithm} 

%% file: numericalresult.tex
\section{Numerical Results}
\label{section:numerical_results}
In this section, we provide numerical results obtained via Monte Carlo simulations to examine the performance of our proposed algorithm for maximizing the RIS-aided mmWave MIMO-OFDM system spectral efficiency. Unless otherwise stated, Table \ref{sysparam} presents the values of the system parameters. In our analysis, we compare the spectral efficiency of the proposed PGA algorithm to that of random RIS phases configuration and to a system without any indirect link or RIS. All three comparisons utilize the blockage model of the direct path introduced earlier with parameters listed in Table \ref{tab:sim_params}
\begin{table}[t]
\renewcommand{\arraystretch}{1}
\caption{System Parameters.}
\label{sysparam}
\centering
\begin{tabular}{rl}
\textbf{Parameter} & \textbf{Value}\\
\hline
$D$ & 200 $m$\\
$l_\text{r}$ & 1.8 $m$\\
$N_{\text{ray}}$ & 10\\
$N_{\text{ray}}$ Direct Link (LOS) & 1\\
$N_{\text{ray}}$ Indirect Link (NLOS) & 10\\
$N_{\text{c}}$ & 8\\
$N_{\text{c}}$ Direct Link (LOS) & 1\\
$N_{\text{c}}$ Indirect Link (NLOS) & 5\\
LOS path loss exponent & 2\\
NLOS path loss exponent & 4\\
Angular spread & $10^{\circ}$\\
Number of subcarriers ($K$) & 24\\
Number of taps ($L_1, L_2, L_3)$ & 3, 4, 5\\
Number of antennas at BS & 64\\
Number of antennas at UE & 4\\
Antenna Gain & 62 dB\\
SNR & -5, 10 dB\\
Number of RIS elements & 64\\
Carrier Frequency & 28 GHz\\
Bandwidth & 850 MHz \\
Initial Learning Rate & .1\\
Convergence Error Threshold & .001\\
Number of Monte Carlo Iterations & 500\\
\end{tabular}
\label{tab:sim_params}
\end{table}

\begin{figure}[t!]
    \centering
    \includegraphics[width=\linewidth]{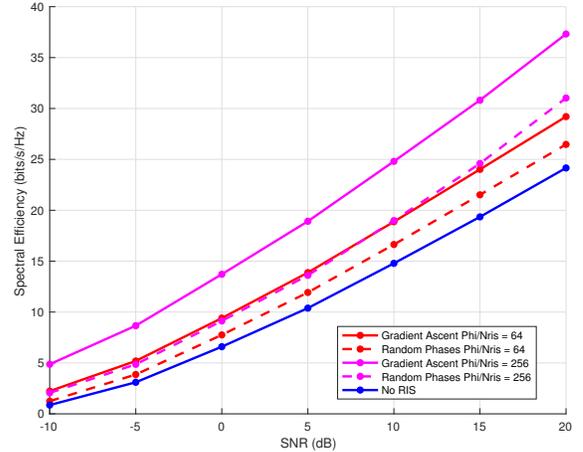}
    \caption{Spectral efficiency versus SNR for two settings of $N_{\text{RIS}}$}
    \label{fig:se_vs_snr}
\end{figure}

First, we consider $l_\text{t} = 10 \text{m}$ and $d_\text{ris} = 2.2 \text{m}$. In Fig. \ref{fig:se_vs_snr}, we observe how the spectral efficiency scales for various SNR settings. The proposed PGA approach achieves substantial spectral efficiency performance gain over a system without a RIS and a RIS-aided system with random RIS phases. As SNR grows, gains of the proposed PGA method and random RIS configuration grow marginally over a system with no RIS. Increasing the number of RIS elements $N_\text{RIS}$, unlocks further spectral efficiency gains as indicated in Fig. \ref{fig:se_vs_snr} for the settings of $N_\text{RIS} = 64$ and $N_\text{RIS} = 256$. We note an interesting tradeoff in which the setting of $N_\text{RIS} = 64$ optimized with the proposed PGA method leads to similar performance to that of the case of random RIS phase configuration for $N_\text{RIS} = 256$.

\begin{figure}[t!]
    \centering
    \includegraphics[width=\linewidth]{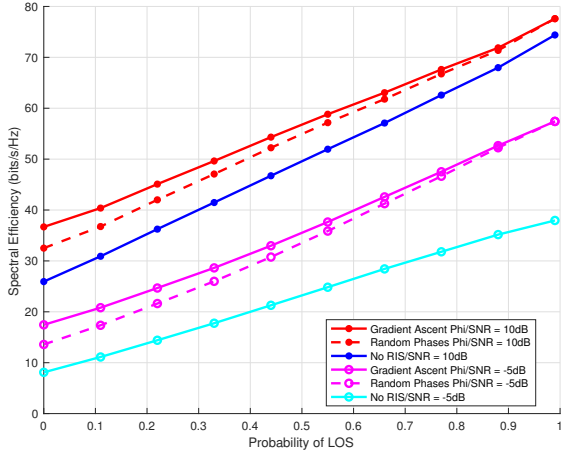}
    \caption{Probability of LOS on the direct path between BS-RIS versus spectral efficiency in the low and high SNR setting}
    \label{fig:plos_vs_se}
\end{figure}

With the incorporation of blockage modeling on the direct path between the BS and UE, we investigated the impact of the probability of the direct path being LOS on the spectral efficiency of the user at SNR settings of $-5$dB and $10$dB. In this experiment, we considered the settings $D = 200 \text{m}$, $l_\text{t} = 5 \text{m}$, $d_\text{ris} = 2.2 \text{m}$. In Fig. \ref{fig:plos_vs_se}, as the probability of LOS increases, blockage on the direct path becomes less likely and spectral efficiency performance of the random RIS phases configuration converges with that of the proposed PGA method due to the presence of a strong direct path between the BS and UE. When the SNR is $10$dB, the spectral efficiency of a system without a RIS approaches that of a RIS-assisted system with proposed PGA approach and random RIS phases configuration, as the direct path experiences becomes less likely to be in blockage. At $\text{SNR}=-5$dB setting, the spectral efficiency is consistently lower than a RIS-assisted system using either the proposed PGA approach or random RIS phase configuration, indicating that the direct path between the BS and UE being available with high probability is not sufficient to compensate for the low SNR in the system. 

\begin{figure}[t!]
    \centering
    \includegraphics[width=\linewidth]{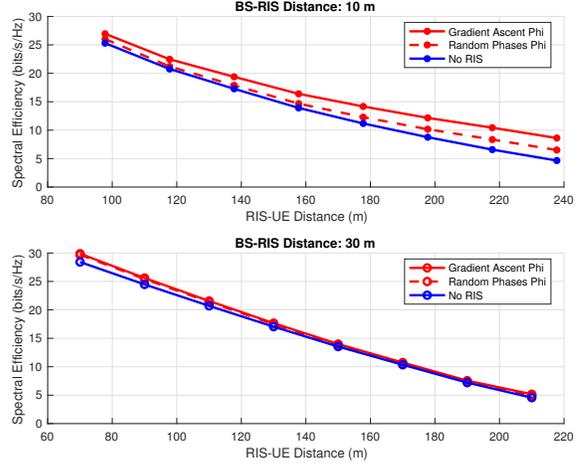}
    \caption{Distance between BS and UE ($d_2$) versus spectral efficiency for two settings of the distance between the BS and RIS ($d_1$)}
    \label{fig:distance_vs_se}
\end{figure}

Lastly, we studied the impact of varying the distances in the system on spectral efficiency. We consider the settings $l_\text{t} = 20 \text{m}$, $d_\text{ris} = 30 \text{m}, \text{SNR} = 5 \text{dB}$, and the distance between the BS and UE $D$ is varied between $100 \text{m}$ and $250 \text{m}$. The distance between the RIS and UE $d_\text{2}$ is evaluated based on $D$ and $d_\text{ris}$ using the expressions in Section \ref{section:system_model}. From Fig. \ref{fig:distance_vs_se}, we observe the RIS needs to be placed in proximity of the BS in order to achieve spectral efficiency gains over that of a system without a RIS. In the case of $d_1 = 10$m and a low setting of $d_2$, $\rho_{\text{direct}}$ and $\rho_{\text{indirect}}$ are lower, which coupled with the high SNR setting, RIS-assisted and no-RIS systems performing similarly. Note this is a similar behavior to the high SNR setting in Fig. \ref{fig:plos_vs_se}. In Fig. \ref{fig:distance_vs_se}, as distance $d_2$ increases both $\rho_{\text{direct}}$ and $\rho_{\text{indirect}}$ increase and the utility of the RIS-assisted system is observed, with the PGA approach outperforming a non-optimized random phase RIS configuration. In the case that $d_1$ is $30$m, the RIS is not sufficiently close to either the BS or the UE in order for the RIS-assisted systems to achieve performance gains over the system without a RIS. The results confirm the RIS should be placed either near to the BS or the UE to avoid the multiplicative effect in pathloss \cite{for_siso_setting}.

%% file: computational_complexity.tex
\section{Computational Complexity}
\label{section:complexity}
In this section, we present the computational complexity of the proposed joint optimization of the RIS phases through the PGA algorithm and power allocation matrices through the spatial-frequency waterfilling algorithm. We provide a Big-O approximation of the complexity of the proposed algorithm, by adopting the complexity analysis presented in \cite{achievable_rate_opt_MIMO}. We then used the Lightspeed matlab toolbox to exemplify how at runtime an estimate of the number of floating-point operations (FLOPs) \cite{lightspeed_toolbox}, evaluation time and the number of iterations scales with $N_\text{RIS}$.

Using rules provided in \cite{achievable_rate_opt_MIMO}, we obtain the Big-O approximation of the computational complexity of the proposed algorithm by evaluating the number of complex-valued multiplications which are required from the primary components of the algorithm. The approximation of the complexity is a result of evaluating the gradient of the spectral efficiency with respect to $\bm\Phi$, the gradient update of $\bm\Phi$, the projection of the elements $\bm\Phi$ back onto the feasible set, the updated evaluation of $\bH_\text{eq}$, and spatial-frequency waterfilling which is employed in the algorithm. In the evaluation of the gradient, we compute $\bH_\text{eq}$ which requires $K(N_\text{r} N_\text{RIS} + N_\text{r}N_\text{t}N_\text{RIS})$ complex-valued multiplications. Similarly, the gradient requires the evaluation of $\bY[k] \forall k$ requiring $K(N_\text{RIS}N_\text{t}^2 + N_\text{r}N_\text{t}N_\text{RIS})$ complex-valued multiplication, the evaluation of $\bZ[k] \forall k$ which requires $K(N_\text{RIS}N_\text{t}^2 + N_\text{RIS}^2N_\text{t} + N_\text{RIS}^2 + N_\text{RIS}^2N_\text{r})$ complex-valued multiplication. Using \cite{achievable_rate_opt_MIMO}, the cost of evaluating $\bA_k(\bm\Phi) \forall k$ is $K(N_\text{t}N_\text{r} + \frac{3}{2}N_\text{t}N_\text{r}^2 + N_\text{r}^3)$. We find the number of complex-valued multiplications for evaluating \eqref{overall_deriv_expression} for each subcarrier as $K(N_\text{r}N_\text{RIS} + N_\text{r}^2N_\text{RIS} + N_\text{r}^3)$. The gradient update requires multiplying the learning rate $\mu$ by the gradient $\nabla_{\mathbf{\Phi}} {\cal R}(\bQ[k],\bm \Phi)$ and the projection onto the feasible set from \eqref{projection} both lead to $N_\text{RIS}$ complex-valued multiplications. Within the spatial-frequency waterfilling method, an SVD of each of the square matrices $\bH_\text{eq}^H[k]\bH_\text{eq}[k] \forall k$ is required. The evaluation of each of these matrices requires $K(N_\text{t}^2N_\text{r})$ complex-valued multiplications. For an $m$ x $n$ matrix, the approximated complexity of and SVD is ${\cal O}(m^2 n + n^3)$. Using this result, for $K$ subcarriers the complexity of the required SVDs is $K{\cal O}(2 N_\text{t}^3)$. We evaluate $\bQ[k] \forall k$ as $\bU[k]\diag(\bp[k])\bU[k]^H$ where $\bU[k]$ is the right singular matrix of $\bH_\text{eq}[k]$ and $\bp[k]$ are the power allocations found by \eqref{power_values} for the $k$-th subcarrier. The evaluation of $\bQ[k] \forall k$ requires $K(2N_\text{RIS}^3)$. Overall, we obtain that approximated complexity of the proposed joint optimization of $\bm\Phi$ and $\bQ[k] \forall k$ as ${\cal O}(K(2N_\text{RIS}^3 + 2N_\text{t}^3 + (N_\text{RIS} + 1)N_\text{r}^3 + (2N_\text{r} + N_\text{r}^2 + N_\text{t} + 1)N_\text{RIS}^2 + (2N_\text{r} + 2N_\text{RIS})N_\text{t}^2 + \frac{3}{2}N_\text{t}N_\text{r}^2 + (2 + 2N_\text{r} + 3N_\text{t}N_\text{r})N_\text{RIS}))$. We note that we can reduce the complexity of the algorithm by considering uniform power allocation over spatial-frequency waterfilling. Full comparison of the two power allocations schemes will be presented in future work.

In order to further inspect the complexity of the proposed algorithm, we use the Lightspeed toolbox \cite{lightspeed_toolbox} to get the number of FLOPs required at runtime. We note that the Lightspeed toolbox provides FLOP counts based on real-valued operations. We update the code to consider both complex-valued multiplications which require 6 FLOPs composed of 4 real multiplications and 2 real summations and complex-valued divisions requiring 8 FLOPs composed of 4 real multiplications, 2 real squarings, 3 real additions and 2 real divisions. In Table \ref{tab:complexity_table}, we provide runtime complexity statistics for varying $N_\text{RIS}$ at $\text{SNR} = -5$dB we observe that as $N_\text{RIS}$ increases the runtime and number of iterations for convergence increases. We observe for $N_\text{RIS} = 324, 400$, the number of FLOPs decreases despite the number of iterations increasing. We observe the same general trends when we run this experiment for $\text{SNR} = 5$dB.

\begin{table}[t!]
 \centering
  \caption{Complexity evaluated by Iteration and FLOP counts for different $N_\text{RIS}$ settings.}
    \begin{tabular}{llll}
		\toprule
		\textbf{$N_\text{RIS}$} & \textbf{Iter. Count} & \textbf{Flop Count} & \textbf{Runtime (s)}  \\
		$4$ & $6$ & 1.70e+10 & 0.1742\\
		$16$ & $8$ & 6.17e+10 & 0.2646\\
		$36$ & $10$ & 1.28e+11 & 0.4617\\
		$144$ & 11 & 1.88e+12 & 1.5274\\
		196 & 12 & 2.28e+12 & 2.9407 \\
		324 & 16 & 6.61e+11 & 12.0429\\
		400 & 17 & 5.6e+11 & 22.8954\\
    \bottomrule
    \end{tabular}%
  \label{tab:complexity_table}%
\end{table}%

%% file: conclusion.tex
\section{Conclusion}
\label{section:conclusion}
In this work, we consider a mmWave frequency-selective setting, in which a PGA algorithm in conjunction with spatial-frequency waterfilling jointly optimizes the RIS phase matrix and power allocation matrices per subcarrier. This work is an extension from the sub-6 GHz narrowband MISO and MIMO work in \cite{pooja_miso,achievable_rate_opt_MIMO}. Unlike prior work, we model URAs at both the BS and UE and also model blockage along the direct path between the BS and UE. We present performance gains of the optimized RIS-assisted system over a non-optimized RIS assisted system and system with no path through the RIS and investigate the impact of blockage and RIS position in RIS-assisted systems on spectral efficiency. Additionally, we run a complexity analysis in which we evaluated the Big-O complexity and measured runtime statistics to understand how complexity of the proposed algorithm scales with larger $N_\text{RIS}$. Immediate future work will consist of studying the impact of imperfect CSI on spectral efficiency under mobility and comparing the performance of our proposed approach to the AO algorithm proposed in \cite{capacity_characterization_MIMO}.